\documentclass[superscriptaddress,aps,prd,nofootinbib,reprint]{revtex4-2}
\pdfoutput=1

\usepackage[T1]{fontenc}
\usepackage{amsmath,amsfonts,amssymb,bm}
\usepackage{graphicx, rotating}
\usepackage{epsfig}
\usepackage{latexsym}
\usepackage{graphicx}
\usepackage{xcolor}
\usepackage[english]{babel}
\usepackage[colorlinks,linktocpage,allcolors=blue]{hyperref}
\begin{document}

\hfill   TU-1202

%%%%%%%%%%%%%%% Title %%%%%%%%%%%%%%%%%%
\title{Can baryon asymmetry be explained by a large initial value 
before inflation?}
%%%%%%%%%%%%%%%%%%%%%%%%%%%%%%%%%%%%%%%%

%%%%%%%%%%%%%% Author %%%%%%%%%%%%%%%%%%
\author{Kai Murai}
\affiliation{Department of Physics, Tohoku University, Sendai, Miyagi 980-8578, Japan}
\author{Fuminobu Takahashi}
\affiliation{Department of Physics, Tohoku University, Sendai, Miyagi 980-8578, Japan}
\author{Masaki Yamada}
\affiliation{Department of Physics, Tohoku University, Sendai, Miyagi 980-8578, Japan}
\affiliation{FRIS, Tohoku University, Sendai, Miyagi 980-8578, Japan}
\author{Wen Yin}
\affiliation{Department of Physics, Tohoku University, Sendai, Miyagi 980-8578, Japan}
%%%%%%%%%%%%%%%%%%%%%%%%%%%%%%%%%%%%%%%%

%%%%%%%%%%%%% Abstract %%%%%%%%%%%%%%%%
\begin{abstract}
    We show that the baryon asymmetry of the Universe cannot be explained by a large initial value before inflation because it inevitably predicts correlated baryon isocurvature perturbations that are already excluded by cosmic microwave background observations. 
    Similar arguments can generally be applied to some models of dark matter.
\end{abstract}
%%%%%%%%%%%%%%%%%%%%%%%%%%%%%%%%%%%%%%%%

\maketitle

%%%%%%%%%%%%%% Section %%%%%%%%%%%%%%%
\section{Introduction}
%%%%%%%%%%%%%%%%%%%%%%%%%%%%%%%%%%%%%%

The origin of the baryon asymmetry of the Universe is a longstanding mystery in cosmology. It is commonly believed that any pre-existing asymmetries are diluted by inflation, and therefore baryon asymmetry should be produced after inflation.
Various models of baryo/leptogenesis have been studied in the literature (see, e.g., Refs.~\cite{Dine:2003ax,Bodeker:2020ghk} for reviews).
However, one can still imagine the possibility that 
a sufficiently large amount of baryon asymmetry is generated before inflation via, e.g., the dynamics of a complex scalar field with $B-L$ charge~\cite{Krnjaic:2016ycc}, so that the amount of baryon asymmetry is consistent with the observed value after the significant dilution by inflation. 

In this letter, we discuss that the scenario of pre-existing baryon asymmetry before inflation predicts correlated baryon isocurvature perturbations and is robustly excluded by the cosmic microwave background (CMB) observations. 
This is because the curvature perturbation is generated independently  by the inflaton fluctuation after the baryon asymmetry is generated. 
Then, we extract the essence and generalize our discussion to show that some
dark matter scenarios are excluded as well.

%%%%%%%%%%%%%% Section %%%%%%%%%%%%%%%
\section{Correlated baryon isocurvature perturbation}
%%%%%%%%%%%%%%%%%%%%%%%%%%%%%%%%%%%%%%

As mentioned above, we assume that the baryon asymmetry is generated before inflation.%
\footnote{More precisely speaking, we assume that the amount of baryon asymmetry in the later universe is determined before inflation. Prior to inflation, it could take any form, such as a large lepton asymmetry or any other type of asymmetry, anything that determines the baryon asymmetry at a later time. Our arguments similarly apply to these cases.} 
After that, the baryon number density, $n_B$, is diluted by the cosmic expansion as
\begin{align}
    n_B 
    \propto 
    a^{-3}
    \ ,
\end{align}
where $a$ is the scale factor.
Although the baryon asymmetry may have initial fluctuations inherent in the generation mechanism, we ignore this because it does not affect our argument and, in fact, only makes it more robust. Below we consider another source of fluctuations.

For clarity, we assume the standard scenario where the curvature perturbation is generated by the fluctuations of the inflaton.
We will discuss other possibilities later. 
The curvature perturbation $\mathcal{R}$ is represented by the fluctuation of the e-folding number $N$ between the flat slicing during inflation and the uniform density slicing during the radiation-dominated era as
\begin{align}
    \mathcal{R}
    =
    \delta N
    \ .
\end{align}
Here the e-folding number is defined by
\begin{align}
    N(t, x) = \int_{t_i}^{t + \delta t(x)}
    H(t') dt',
\end{align}
where $t_i$ is an initial time before the CMB scales exited the horizon, $H$ is the Hubble parameter, and $\delta t$ is the fluctuation of the time on the uniform density slicing.
Note that $\delta t$ can arise from the fluctuation of the duration of both the inflationary and radiation-dominated eras.
Considering $n_B \propto a^{-3} \propto e^{-3N}$, we obtain the fluctuations of the baryon number density on the uniform density slicing as
\begin{align}
    \delta n_B
    =
    - 3 \delta N \bar{n}_B
    \ .
\end{align}
Here and hereafter, we denote unperturbed quantities with bars.
After the QCD phase transition, baryons become nucleons such as  protons and neutrons.
Then, the fluctuation of the non-relativistic baryon energy density, $\rho_B$, is evaluated as
\begin{align}
    \frac{\delta \rho_B}{\bar{\rho}_B} 
    =
    - 3 \mathcal{R}
    \ .
\end{align}
This fluctuation corresponds to the baryon isocurvature mode given by
\begin{align}
    \mathcal{S}_B
    \equiv 
    \frac{\delta \rho_B}{\bar{\rho}_B} 
    -
    \frac{3}{4}\frac{\delta \rho_\gamma}{\bar{\rho}_\gamma} 
    \simeq 
    - 3 \mathcal{R}
    \ ,
    \label{eq: baryon isocurvature result}
\end{align}
where we used $\delta \rho_\gamma \simeq 0$ on the uniform density slicing during the radiation-dominated era.
Note that $\mathcal{S}_B$ is fully anti-correlated with the curvature perturbation.

The isocurvature perturbation is often parameterized by 
\begin{align}
    \beta_\mathrm{iso}(k)
    \equiv 
    \frac{\langle |\mathcal{S}_\mathrm{eff}(k)|^2 \rangle}{\langle |\mathcal{R}(k)|^2 \rangle + \langle |\mathcal{S}_\mathrm{eff}(k)|^2 \rangle}
    \ ,
\end{align}
where the quantities with a comoving wavenumber $k$ represent the Fourier modes.
Here, $\mathcal{S}_\mathrm{eff}$ is an effective matter density isocurvature perturbation translated to the cold dark matter (CDM) perturbations defined by
\begin{align}
    \mathcal{S}_\mathrm{eff}
    \equiv 
     \mathcal{S}_\mathrm{CDM} + 
    \frac{\Omega_B}{\Omega_\mathrm{CDM}}
    \mathcal{S}_B
    \ ,
    \label{eq: Seff def}
\end{align}
where $\mathcal{S}_\mathrm{CDM}$ is the CDM isocurvature perturbation similarly defined as Eq.~\eqref{eq: baryon isocurvature result}, and $\Omega_B$ and $\Omega_\mathrm{CDM}$ are the density parameter of the baryon and CDM, respectively.

The matter density isocurvature perturbation that is fully anti-correlated and shares the same spectral tilt with the curvature perturbation is constrained as~\cite{Planck:2018jri}
\begin{align}
    \beta_\mathrm{iso}(k) < 10^{-3}
    \ ,
\end{align}
at $k = 0.002\,\mathrm{Mpc}^{-1}$, $0.05\,\mathrm{Mpc}^{-1}$, and $0.1\,\mathrm{Mpc}^{-1}$.
Now we consider the baryon isocurvature mode. Then, we obtain the constraint on this mode as
\begin{align}
    \left\vert \mathcal{S}_B(k) \right\vert
    <
    \frac{\Omega_\mathrm{CDM}}{\Omega_B} \sqrt{\frac{10^{-3}}{1 - 10^{-3}}}\mathcal{R}(k)
    \simeq 
    0.17 \mathcal{R}(k)
    \ ,
\end{align}
where we used $\Omega_B h^2 = 0.022$ and $\Omega_\mathrm{CDM} h^2 = 0.12$~\cite{Planck:2018vyg} with $h$ being the reduced Hubble constant.
From the combination of the CMB and large-scale structure observations, a similar constraint is obtained~\cite{Seljak:2006bg}.
Thus, scenarios that generate large baryon asymmetry prior to inflation are  excluded from observations, as they result in fully anti-correlated baryon isocurvature fluctuations given in Eq.~\eqref{eq: baryon isocurvature result}.%
\footnote{%
If the CDM has fully correlated isocurvature perturbations that cancel the baryon isocurvature perturbations in Eq.~\eqref{eq: Seff def}, our argument can be evaded.
Such CDM isocurvature perturbations 
can be generated by the mechanism discussed in Sec.~\ref{sec:CDM} if $(\partial \ln \rho_{\rm CDM} / \partial X_*) (\dot{X}_* / H_*) \simeq 3 \Omega_{\rm B} / \Omega_{\rm CDM}$ ($\simeq 3/5$) (see Eq.~(\ref{eq:SCDM})). 
A scenario with a similar cancellation has been discussed in Ref.~\cite{Harigaya:2019uhf} in a different context for baryogenesis after inflation. 
}
If the baryon asymmetry had the initial fluctuations, it would lead to independent baryon isocurvature perturbations, further strengthening the argument.

%%%%%%%%%%%%%% Section %%%%%%%%%%%%%%%
\section{Correlated DM isocurvature perturbation}
%%%%%%%%%%%%%%%%%%%%%%%%%%%%%%%%%%%%%%
\label{sec:CDM}

Here we extend the discussion to the case with other pre-existing components in the Universe based on the $\delta N$ formalism~\cite{Starobinsky:1985ibc,Salopek:1990jq,Sasaki:1995aw}.
For concreteness, we specifically consider CDM.
Let us consider the dark matter energy density, $\rho_{\mathrm{CDM}}$, at a time when the density is already fixed after inflation, and the CMB scale is still superhorizon.
Suppose that $\rho_{\mathrm{CDM}}$ is a function of some quantity $X$ other than inflaton field value $\phi$, especially at the time when the CMB scale exits the horizon during inflation.
In the following, we focus on the fluctuations on the CMB scale and denote quantities at the horizon exit of the CMB scale with a subscript $*$.
We thus assume $\rho_{\rm CDM} = \rho_{\rm CDM} (\phi_*, X_*)$. 
The parameter $X_*$ depends on the model and is not specified in our argument below. It may be identified as a field value of light boson DM, as we will discuss shortly.

The CDM isocurvature can be written as
\begin{align}
    \mathcal{S}_\mathrm{CDM}
    =
    3(\zeta_\mathrm{CDM} - \zeta_\gamma)
    \ ,
\end{align}
where $\zeta_\mathrm{CDM}$ and $\zeta_\gamma$ are curvature perturbations on the uniform density slicing with respect to CDM and photon, respectively.

Since the fluctuation of the photon energy density originates from the inflaton fluctuation, $\zeta_\gamma$ is given by
\begin{align}
    \zeta_\gamma
    &=
    \frac{\mathrm{d} N_\gamma}{\mathrm{d} \phi_*} \delta \phi_*
    = 
    - H_* \frac{\delta \phi_*}{\dot{\phi}_*}
    =
    \mathcal{R}
    \ .
\end{align}
Here, the e-folding number $N_\gamma \simeq N$ is evaluated between the flat slicing during inflation and the uniform density slicing for photon during the radiation-dominated era.

On the other hand, since we assume that the CDM density depends on the parameter $X_*$ in addition to the inflaton field value $\phi_*$, the uniform density slicing for CDM is different from that for photon, and $\zeta_\mathrm{CDM}$ receives other contributions. 
Since $\zeta_\mathrm{CDM}$ is the curvature perturbations when $\phi_*$ and $X_*$ receive fluctuations, it is given by
\begin{align}
    \zeta_\mathrm{CDM} 
    =
    & N_c(\bar{\phi}_* + \delta \phi_*, \bar{X}_*(\bar{\phi}_* ) + \delta X_*)    
    \nonumber \\
    & - N_c(\bar{\phi}_*, \bar{X}_*(\bar{\phi}_*))    
    \ ,
    \label{eq: zeta_CDM diff in N}
\end{align}
where the e-folding number $N_c$ is now evaluated between the flat slicing during inflation and the uniform density slicing for $\rho_{\rm CDM}(\phi_*, X_*)$. 
We explicitly write the time dependence of $\bar{X}_*$ via $\bar{\phi}_*$-dependence by regarding $\bar{\phi}_*$ as a timer field.
Thus, $\zeta_\mathrm{CDM}$ is written in terms of derivatives as
\begin{align}
    \zeta_\mathrm{CDM}
    &=
    \frac{\partial N_c}{\partial \phi_*} \delta \phi_*
    +
    \frac{\partial N_c}{\partial X_*}
    \delta X_* 
    \nonumber \\
    &=
    \frac{\mathrm{d} N_c}{\mathrm{d} \phi_*} \delta \phi_*
    - \frac{\partial N_c}{\partial X_*}
    \dot{X}_* \frac{\delta \phi_*}{\dot{\phi}_*}
    +
    \frac{\partial N_c}{\partial X_*}
    \delta X_*
    \ , 
    \label{eq: zeta_CDM formula}
\end{align}
where $\delta \phi_*$ and $\delta X_*$ are evaluated on the flat slicing.
Since the inflaton can be regarded as a timer field, we can consider the first term in the second line to denote the fluctuation of the time, and therefore this term should be identified with the ordinary curvature perturbation, $\zeta_\gamma$ or $\mathcal{R}$.
The second term comes from the difference between the total derivative and partial derivative, which is a new source of isocurvature perturbations for pre-existing DM.
The last term expresses the isocurvature perturbation due to fluctuations of the parameter itself that arise independently of the inflaton fluctuations, which has been discussed widely and we do not focus on in this letter. 

The second term in the right-hand side of Eq.~(\ref{eq: zeta_CDM formula}) gives isocurvature perturbation, 
\begin{align}
    \mathcal{S}_\mathrm{CDM}
    \simeq \frac{\partial \ln \rho_{\rm CDM}}{\partial X_*}
     \frac{\dot{X}_*}{H_*} \mathcal{R}
    \ ,
    \label{eq:SCDM}
\end{align}
where we use $\partial N_c / \partial X_* = (1/3) \partial \ln \rho_{\rm CDM} / \partial X_*$. 
We thus obtain a constraint,
\begin{align}
 \left\vert \frac{\partial \ln \rho_{\rm CDM}}{\partial X_*}
     \frac{\dot{X}_*}{H_*} \right\vert 
 \; \lesssim \;
 0.032.
\end{align}
If this is not satisfied, pre-existing dark matter is excluded by the CMB observations in a similar way to the pre-existing baryon asymmetry.
One example of excluded scenarios is a light scalar dark matter moving before the CMB scale exits the horizon~\cite{Caputo:2023ikd}, where $X_*$ is identified as the scalar field value.
Another example is the misalignment production of hidden photon dark matter with an exponentially large initial field value, where $X_*$ is identified as the amplitude of the hidden photon, $A_i \propto a^{-1}$. (See Refs.\,\cite{Nelson:2011sf,Arias:2012az,Nakayama:2019rhg,Nakayama:2020rka,Kitajima:2023fun} for the efforts in model-buildings against the suppression of $a$.)

We note that the standard misalignment mechanism~\cite{Preskill:1982cy,Abbott:1982af,Dine:1982ah} for a scalar field does not suffer from this type of isocurvature perturbations if its mass is much smaller than $H_*$. 
In this case, one may identify $X_*$ as the field value of the scalar field, which is almost constant during inflation, i.e., $\dot{X_*} \simeq 0$ 
and the second term of Eq.~(\ref{eq: zeta_CDM formula}) is negligibly small. 
Then the last term of Eq.~(\ref{eq: zeta_CDM formula}) can be important 
because it gives $\delta X_* \simeq H_* / (2\pi)$. 
This contribution has been extensively discussed in the literature~\cite{Lyth:1991ub,Kobayashi:2013nva}.
 A similar discussion can be applied to baryon number density~\cite{Enqvist:1998pf,Enqvist:1999hv,Kawasaki:2001in}.

The argument in this section can be applied to baryon isocurvature perturbations to reproduce Eq.~(\ref{eq: baryon isocurvature result}), where one can identify $X_*$ as the baryon number density. 
This formulation clarifies a possible loophole in our argument. 
If the baryon asymmetry is stored by the inflaton itself and is given by a function solely of the inflaton, 
the second and third terms of Eq.~(\ref{eq: zeta_CDM formula}) are absent, and the isocurvature perturbations are not generated. 

%%%%%%%%%%%%%% Section %%%%%%%%%%%%%%%
\section{Discussions}
%%%%%%%%%%%%%%%%%%%%%%%%%%%%%%%%%%%%%%

We have shown that if the baryon asymmetry is generated before inflation, the fluctuation of the duration of inflation induces baryon isocurvature perturbations proportional to the curvature perturbation at the end of inflation.
As a result, we conclude that the baryon asymmetry of the Universe cannot be explained by large initial values before inflation.
It is worth noting that our argument is unlikely to be avoided by the anthropic argument because galaxies will form even in universes with sizable isocurvature perturbations. 

Note that our result does not exclude baryogenesis during inflation if baryon asymmetry is generated much after the CMB scales exit the horizon. 
Since the inflaton can be identified as a timer field, one can consider a scenario in which baryogenesis is triggered 
at a certain field value of the inflaton
and baryon density is uniformly generated on the comoving slice.
Then, the generated baryon asymmetry has isocurvature perturbations only on smaller scales than the horizon scale at baryogenesis.
Although the baryon isocurvature is also constrained by the inhomogeneous big bang nucleosynthesis on smaller scales than the CMB scales~\cite{Inomata:2018htm}, this constraint does not exclude the baryon isocurvature perturbations of the same order as the curvature perturbations unless the curvature perturbation is significantly enhanced on small scales. 
We, therefore, conclude that baryogenesis must take place
after the CMB scales leave the horizon during inflation.

We emphasize that our discussion can be applied to a component other than the baryon asymmetry such as CDM as long as it exists before the CMB scales exit the horizon during inflation and evolves in time during inflation so that the duration of inflation affects its density in the later universe.
The magnitude of the isocurvature perturbations depends on how the time evolution during inflation affects the density in the later universe. 

Lastly, we would like to mention the similarity between our argument and the generation of baryon asymmetry and/or dark matter in scenarios such as the curvaton scenario and similar ones. It is widely recognized that unless the baryon and/or dark matter is generated after the adiabatic density perturbation is formed by the curvaton, correlated isocurvature perturbations are produced~\cite{Lyth:2002my,Lyth:2003ip,Kitajima:2017fiy}. However, to the best of our knowledge, it has not been recognized that the same argument applies to the standard inflationary scenario as well. The purpose of this letter is to clarify this point and to show definitively that preparing a large initial baryon asymmetry before inflation to account for the observed baryon asymmetry in our Universe is already observationally excluded.

%%%%%%%%%%%%%%%%%%%%%%%%%%%%%%%%%%%%%%
\section*{Acknowledgments}
The present work is supported by JSPS KAKENHI Grant Numbers 20H01894 (F.T.), 20H05851 (F.T., M.Y., and W.Y.), 21K20364 (W.Y.), 22H01215 (W.Y.), 22K14029 (W.Y.), 23K13092 (M.Y.), and 23KJ0088 (K.M.), and JSPS Core-to-Core Program (grant number: JPJSCCA20200002) (F.T.).
MY was supported by MEXT Leading Initiative for Excellent Young Researchers. 
This article is based upon work from COST Action COSMIC WISPers CA21106, supported by COST (European Cooperation in Science and Technology).
%%%%%%%%%%%%%%%%%%%%%%%%%%%%%%%%%%%%%%

%%%%%%%%%%%% References %%%%%%%%%%%%%%
\bibliographystyle{utphys}
\bibliography{ref}
%%%%%%%%%%%%%%%%%%%%%%%%%%%%%%%%%%%%%%

\end{document}